\documentclass[cameraready]{Interspeech}

\title{BEST-RQ-2: Contextualize--Then--Predict, a Two-Step Approach for Self-Supervised Audio Representations}

\author[affiliation={1}, orcid=0009-0005-5044-5301, correspondingauthor]{Ludovic}{TUNCAY}
\author[affiliation={1}, orcid=0000-0002-7219-5463]{\'Etienne}{LABB\'E}
\author[affiliation={1}, orcid=0000-0001-8984-1399]{Thomas}{PELLEGRINI}

\address{
  $^1$ IRIT, Universit\'e de Toulouse, CNRS, Toulouse INP, Toulouse, France
}

\email{ludovic.tuncay@irit.fr, etienne.labbe@irit.fr, thomas.pellegrini@irit.fr}


\usepackage{comment}
\usepackage{makecell}
\usepackage[table]{xcolor}
\usepackage{hyperref}
\usepackage{multirow}
\usepackage{booktabs}
\DeclareMathOperator*{\argmin}{arg\,min}
\DeclareMathOperator*{\argmax}{arg\,max}

\begin{document}

\maketitle

\begin{abstract} 
  Self-supervised learning enables audio representations that transfer across domains and tasks. We present \mbox{BEST-RQ-2}, an evolution of BEST-RQ that retains frozen random-projection-based discrete targets while introducing a two-step \emph{contextualize--then--predict} pretraining scheme. A ViT context encoder processes only the unmasked spectrogram regions, and a lightweight predictor infers targets for the masked regions; the predictor is discarded after pretraining.
  Replacing the original Conformer encoder with a ViT shifts performance across domains, slightly reducing speech performance while improving music and environmental sounds, with comparable average scores. The main improvement comes from decomposing masked prediction into separate contextualization and prediction stages. On the X-ARES and XARES-LLM benchmarks, BEST-RQ-2 consistently outperforms one-stage baselines in overall transfer while keeping inference compute unchanged. Code\footnote{\url{https://github.com/LudovicTuncay/audio-embeddings}} and model checkpoints\footnote{\url{https://huggingface.co/ltuncay/BEST-RQ-2}} are publicly available.
\end{abstract}

\section{Introduction}

Audio representations are increasingly expected to transfer across speech, music, and environmental sound while supporting diverse downstream uses, from probing evaluations to front-ends for audio language models. Benchmarks such as X-ARES and XARES-LLM explicitly measure this cross-domain transfer under complementary evaluation protocols~\cite{zhang2025xares, xares-llm}.

Masked prediction with discrete targets has proven effective for self-supervised audio learning. BEST-RQ is particularly attractive because it relies on fixed random-projection quantization, enabling stable cross-entropy training without learned codebooks~\cite{chiu2022self, whetten2024open}. In this work, we revisit the speech-centric BEST-RQ with architectures better matched to generic audio. Replacing the original Conformer~\cite{gulati2020conformer} encoder with a Vision Transformer (ViT) requires spectrogram patch tokenization~\cite{dosovitskiy2020image, gong2022ssast, chen2023beats,baevski2023efficient}, which alters the balance between temporal detail and local time--frequency texture modeling. This redistributes performance across domains while leaving overall transfer performance largely unchanged.

We introduce BEST-RQ-2, which retains BEST-RQ's frozen discrete targets while adopting a two-step contextualize--then--predict decomposition. A ViT context encoder processes only unmasked patches, and a lightweight predictor reconstructs and predicts masked targets during pretraining before being discarded at inference time. To isolate architectural effects, we also introduce BEST-RQ (ViT), a one-stage ViT variant that shares the same tokenizer and targets as BEST-RQ-2.

Our contributions are fourfold. (i) We revisit BEST-RQ using a ViT encoder and show that the resulting model achieves comparable overall transfer while redistributing performance across audio domains. (ii) We introduce a two-step decomposition separating contextualization from prediction. (iii) We demonstrate on X-ARES and XARES-LLM that most performance gains arise from this decomposition rather than from tokenization changes, while inference cost remains unchanged. (iv) We release code\footnotemark[1] and checkpoints\footnotemark[2] for BEST-RQ-2.

\begin{figure*}[t]
  \centering
  \includegraphics[width=0.8\textwidth]{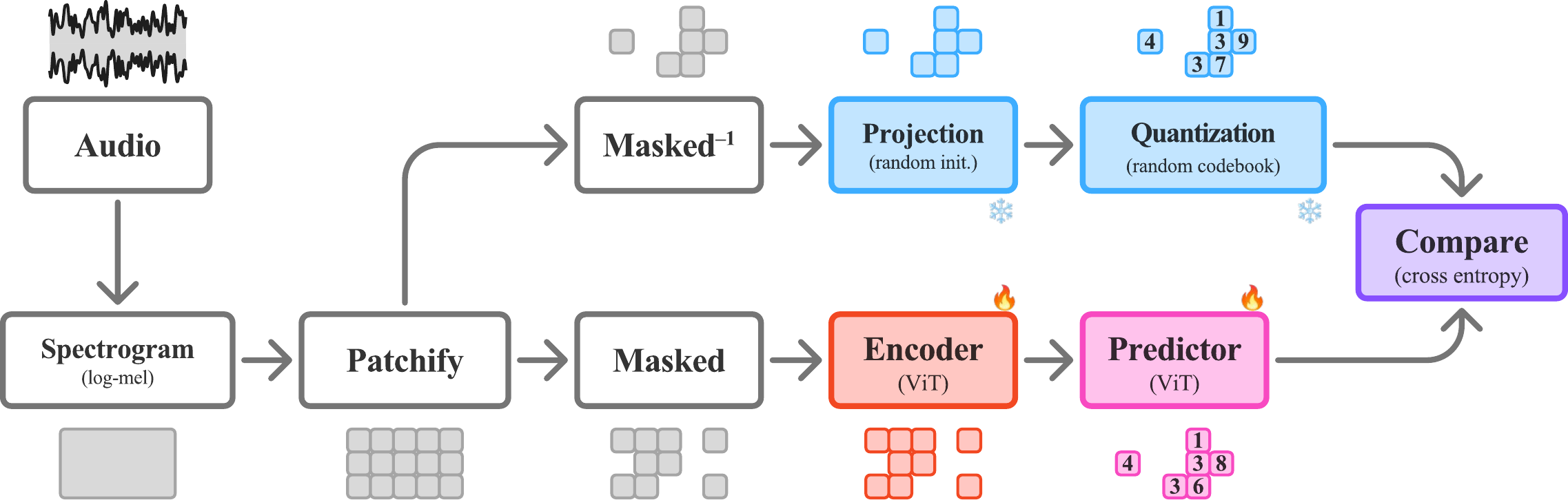}
  \caption{BEST-RQ-2 pretraining pipeline. A log-mel spectrogram is patchified (partitioned into patches) and masked. The ViT context encoder processes unmasked patches and a lightweight ViT predictor outputs logits over $K=8192$ codes for masked patches. Targets are discrete indices generated by a frozen random-projection quantizer. Training minimizes cross-entropy on masked positions.}
  \label{fig:BEST-RQ-2_arch}
  \vspace{-0.2cm}
\end{figure*}

\section{Proposed Method: BEST-RQ-2}
BEST-RQ-2 is a masked prediction method that learns audio representations from log-mel spectrograms using discrete targets from a frozen random-projection quantizer, as in BEST-RQ~\cite{chiu2022self, whetten2024open}. It introduces two changes: (i) a ViT-based architecture and (ii) a two-step encoder--predictor decomposition inspired by Audio-JEPA~\cite{tuncay2025audio}. Figure~\ref{fig:BEST-RQ-2_arch} summarizes the pipeline.

\subsection{Design rationale}
BEST-RQ-2 modifies BEST-RQ along two dimensions: the encoder architecture and the prediction process. The original BEST-RQ encoder is Conformer-based~\cite{gulati2020conformer} and includes convolutional modules with local receptive fields defined over a dense time axis. In BEST-RQ, masked regions are therefore provided \emph{in place} at the encoder input (rather than removed), so that convolutional neighborhoods remain well-defined and the ordering of local patterns is preserved throughout the forward pass.

An encoder--predictor decomposition, where masked tokens are dropped in the encoder and reintroduced later, is not directly compatible with this Conformer design without architectural changes: removing patches breaks the dense layout assumed by convolutions and alters their effective neighborhoods. ViT encoders, in contrast, can operate on an arbitrary subset of patches because self-attention is informed by positional embeddings; masked patches can be omitted in the encoder and reintroduced in a predictor using their positional indices. We therefore adopt a ViT backbone for BEST-RQ-2 to enable the encoder stage to be applied to unmasked regions only.

To disentangle effects, we also introduce BEST-RQ (ViT), which replaces the Conformer encoder with a ViT but retains the \emph{same masking interface as BEST-RQ}: the encoder directly receives masked patches in place. BEST-RQ (ViT) can thus be viewed either as BEST-RQ with a ViT encoder, or as BEST-RQ-2 without the encoder--predictor decomposition.

\subsection{Patch tokenization}
\label{subsec:patch-tokenization}
Given a log-mel spectrogram $X \in \mathbb{R}^{F \times T}$, we partition it into non-overlapping $P\times P$ patches ($P{=}16$), yielding $N=(F/P)(T/P)$ tokens. Each patch $x_i \in \mathbb{R}^{P^2}$ is mapped to an embedding $e_i \in \mathbb{R}^{D}$ via a learned linear projection, as in ViT-style spectrogram encoders~\cite{dosovitskiy2020image, gong2022ssast, chen2023beats,baevski2023efficient}.

\subsection{Masking}
We sample a mask ratio $r$ and select masked indices $M$ uniformly without replacement over the 2D patch grid, following common masked-prediction audio pretraining setups~\cite{huang2022masked, baade2022mae, baevski2023efficient, niizumi2023masked}. We set $|M|=\lfloor rN \rfloor$ and define the unmasked set as $V=\{1,\ldots,N\}\setminus M$.

In the original BEST-RQ, masked regions are replaced by random noise and provided in place to the encoder. For transformer-based variants (BEST-RQ (ViT) and BEST-RQ-2), masked patches are represented using a learned mask token. BEST-RQ (ViT) keeps in-place masking, while BEST-RQ-2 removes masked patches from the encoder input and reintroduces them in the predictor with added positional embeddings. The loss is computed only on masked positions.

\subsection{Two-step encoder--predictor}
Similar to JEPA-style architectures~\cite{assran2023self, fei2023jepa, niizumi2024masked, riou2024stem, riou2024investigating, tuncay2025audio, assran2025v}, BEST-RQ-2 factorizes masked prediction into a context encoder $f_{\theta}$ and a predictor $g_{\phi}$. The context encoder (ViT) processes only unmasked patch embeddings (with positional embeddings) and outputs contextualized representations:
\begin{equation}
  \{h_i\}_{i\in V} = f_{\theta}\big(\{e_i\}_{i\in V}\big).
\end{equation}
The predictor receives (i) $\{h_i\}_{i\in V}$ and (ii) learned mask embeddings at $i\in M$ (with positional embeddings), and produces masked representations:
\begin{equation}
  \{z_i\}_{i\in M} = g_{\phi}\big(\{h_i\}_{i\in V}, \{\text{mask}\}_{i\in M}\big).
\end{equation}
A linear classifier maps $z_i$ to logits over $K$ codes. The predictor is used only during pretraining and discarded at inference time.

\subsection{Frozen random-projection targets}
As in BEST-RQ~\cite{chiu2022self, whetten2024open}, each patch $x_i$ is mapped to a discrete target via a frozen random projection followed by nearest-code assignment. The random projection matrix $R$ is sampled once using a Xavier/Glorot uniform initialization~\cite{glorot2010understanding}. We compute
\begin{equation}
  u_i = R x_i, \qquad R \in \mathbb{R}^{d \times P^2},
\end{equation}
and assign the nearest code in a fixed codebook $C=\{c_1,\ldots,c_K\}$ with $c_k \in \mathcal{S}^d$, the hypersphere of dimension $d$:
\begin{equation}
  y_i = \argmin_{k \in \{1,\ldots,K\}} \lVert u_i - c_k \rVert_2 = \argmax_{k \in \{1,\ldots,K\}} \, u_i^\top c_k,
\end{equation}
yielding $y_i \in \{1,\ldots,K\}$. The projection matrix $R$ and codebook $C$ are sampled once and kept frozen during training.

\subsection{Training objective}
Let $p_{i}(k \mid X_V)$ denote the softmax probability of code $k$ predicted for a masked patch $i \in M$, given unmasked input $X_V$. We minimize cross-entropy over masked positions:
\begin{equation}
  \mathcal{L} = \frac{1}{|M|} \sum_{i \in M} \mathrm{CE}\big(p_{i}(\cdot \mid X_V), y_i\big).
\end{equation}

\section{Implementation Details}
\subsection{Audio preprocessing}
For BEST-RQ-2, BEST-RQ (ViT) and Audio-JEPA, audio is resampled to 16~kHz mono and converted to a 128-bin log-mel spectrogram using a 128~ms analysis window and a 39.0625~ms hop, matching Audio-JEPA preprocessing and yielding $T{=}256$ frames for a 10~s clip. The spectrogram $X\in\mathbb{R}^{128\times256}$ is partitioned into non-overlapping $16\times16$ patches, giving $N=(128/16)(256/16)=128$ tokens. During pretraining, we sample $r\sim\mathcal{U}[0.4,0.6]$ and mask exactly $\lfloor rN\rfloor$ patches; the loss is computed only on masked positions.

\subsection{Models and training setup}
BEST-RQ follows the published recipe and open implementation of~\cite{whetten2024open}. BEST-RQ-2 and BEST-RQ (ViT) use the same frozen random-projection quantizer as BEST-RQ with \mbox{$K=8192$} codes and projection dimension $d{=}16$~\cite{chiu2022self, whetten2024open}. The context encoder is a ViT with embedding dimension 768, depth 12, 12 attention heads, MLP ratio 4.0, drop-path 0.1, and 2D sine--cosine positional embeddings. BEST-RQ-2 uses an additional lightweight ViT predictor with the same embedding and attention configuration but depth 4; it is discarded after pretraining, so inference uses only the encoder. BEST-RQ (ViT) is the one-stage ablation that uses the same patchification, masking, and targets as BEST-RQ-2 but predicts masked-patch logits with a single ViT (no separate predictor).

\subsection{Optimization and compute}
\label{subsec:optim}
BEST-RQ-2 and BEST-RQ (ViT) are trained with AdamW using a learning rate of $1\times10^{-4}$, weight decay 0.05, and linear warmup over the first 5\% of training (10k steps), followed by cosine decay, for a total of 200k optimization steps in fp32. BEST-RQ is trained for the same number of steps using the optimizer, schedule, and fp16 configuration from the open implementation of~\cite{whetten2024open}. Models are matched in dataset split and number of optimization steps, but not strictly in the number of examples processed due to differing recommended batch sizes. All runs are executed on a single NVIDIA H100 GPU.

To contextualize deployment cost, Table~\ref{tab:compute} reports parameter counts and forward-pass compute (GFLOPs) for a 10~s input, separately for training (including on-the-fly random-projection target computation) and inference (encoder only). We also report relative inference speed measured on our hardware under a fixed batch size and normalized to BEST-RQ-2; note that BEST-RQ uses fp16 while ViT-based models use fp32.

\begin{table}[t]
  \centering
  \caption{Model size and compute for a 10~s input. Inference (Inf.) uses the encoder only. Relative speed is measured on a single NVIDIA H100 and normalized to BEST-RQ-2.}
  \vspace{-0.1cm}
  \label{tab:compute}
  \setlength{\tabcolsep}{3pt}
  \rowcolors{2}{}{gray!13}
  \begin{tabular}{lccccc}
    \hline
    & \multicolumn{2}{c}{\textbf{Params (M)}} & \multicolumn{2}{c}{\textbf{GFLOPs}} & \textbf{Rel.\ speed} \\
    \cmidrule(lr){2-3} \cmidrule(lr){4-5} \cmidrule(lr){6-6}
    \multirow{-2}{*}{\textbf{Model}} & \textbf{Train} & \textbf{Inf.} & \textbf{Train} & \textbf{Inf.} & \textbf{Inf.} \\
    \hline
    BEST-RQ & 83 & 83 & 49.0 & 46.5 & 0.97$\times$ \\
    BEST-RQ (ViT) & 92 & 85 & 23.4 & 22.5 & 1.0$\times$ \\
    BEST-RQ-2 & 120 & 85 & 30.1 & 22.5 & 1.0$\times$ \\
    \hline
  \end{tabular}
  \vspace{-0.2cm}
\end{table}

Although BEST-RQ has roughly twice the inference GFLOPs, its fp16 precision provides nearly a $2\times$ throughput advantage on H100 hardware, resulting in comparable inference speed to the fp32 ViT-based models. BEST-RQ-2 adds parameters only during training via the predictor, which is discarded at inference, leaving inference cost identical to BEST-RQ (ViT).

\section{Experimental Setup}

\subsection{Pretraining data}
All models are pretrained on AudioSet~\cite{gemmeke_audio_2017}. After identical preprocessing and filtering (e.g., removal of silent or corrupted clips), the training split contains approximately 1.9M 10~s clips spanning speech, music, and environmental sounds. BEST-RQ, BEST-RQ (ViT), and BEST-RQ-2 are trained on this same split to isolate architectural effects.

\subsection{Evaluation benchmarks}
We evaluate representations using two complementary benchmark suites.
\begin{enumerate}
  \item \textbf{X-ARES~\cite{zhang2025xares}.}
    X-ARES evaluates general-purpose audio encoders across speech, environmental sound, and music using both linear probing (MLP) and non-parametric kNN protocols. We follow the official evaluation pipeline and report domain-averaged metrics. Results are reported on 21 datasets due to an issue affecting one dataset in the official release.
  \item \textbf{XARES-LLM~\cite{xares-llm}.}
    XARES-LLM evaluates frozen encoders used as front-end modules for large audio language models within the AECC 2026 pipeline. The benchmark measures downstream classification and understanding performance after LALM training, providing a complementary evaluation of encoder usefulness beyond probing.
\end{enumerate}

\subsection{Training protocol and fairness}
All BEST-RQ variants are pretrained for 200k optimization steps on the same AudioSet split under a comparable compute budget ($\sim$17--18~h on a single NVIDIA H100 GPU). For BEST-RQ, we follow the implementation-recommended batch size and FP16 training setup. For BEST-RQ (ViT) and BEST-RQ-2, we use the parameters detailed in Subsection~\ref{subsec:optim}. Comparisons are therefore controlled for data split, optimization steps, and compute budget, but not strictly for the total number of examples processed. As the closest two-step contextualize--then--predict baseline, we chose to train Audio-JEPA from scratch on the same AudioSet split, with identical preprocessing, the same 200k optimization steps, and a matched training budget ($\sim$18~h). For other hyperparameters, we follow the open implementation~\cite{tuncay2025audio}.

\section{Results}
\label{sec:results}

All encoders are evaluated frozen and output a \emph{sequence} of token embeddings, as required by the X-ARES and XARES-LLM pipelines. We extract token representations from the final encoder block at each model's native tokenization (patches for BEST-RQ (ViT) and BEST-RQ-2; strips for BEST-RQ), with no additional pooling or temporal aggregation beyond what each benchmark applies. We report domain-averaged scores (Speech/Environment/Music) for both benchmarks and primarily use MoM (mean-of-means), defined as the unweighted average of the three domain means, because it equally weights Speech, Environmental sound, and Music regardless of how many datasets each domain contains. We also report the dataset-level Overall mean for completeness.

Because X-ARES and XARES-LLM use different dataset suites and evaluation pipelines, results should be interpreted within each benchmark and are not directly comparable across them.

As noted earlier, we trained BEST-RQ variants and Audio-JEPA under our controlled pretraining protocol. We also re-evaluated all baseline encoders using the official evaluation pipelines to ensure comparability.

\subsection{X-ARES}

\subsubsection{Linear Probing}

\begin{table}[t]
  \centering
  \setlength{\tabcolsep}{3pt}
  \caption{Condensed X-ARES linear-probing results ($\uparrow$), averaged by audio domain. Bold and underlined values indicate the best and second-best score, respectively, within each column across the compared models. Env.\ denotes the environmental-sound domain. MoM is the mean of the three domain means; Overall is the mean over all datasets.}
  \label{tab:xares-mlp}
  \rowcolors{2}{white}{gray!13}
  \begin{tabular}{lccccc}
    \hline
    \textbf{Model} & \textbf{Speech} & \textbf{Env.} & \textbf{Music} & \textbf{MoM} & \textbf{Overall} \\
    \hline
    data2vec~\cite{baevski2022data2vec} & \underline{.66} & .17 & .31 & .38 & .43 \\
    wav2vec~2.0~\cite{baevski2020wav2vec} & .63 & .29 & .43 & .45 & .48 \\
    Whisper~\cite{radford2023robust} & \textbf{.73} & .29 & .44 & \underline{.49} & \textbf{.53} \\
    Audio-JEPA & .43 & .31 & .52 & .42 & .41 \\
    BEST-RQ & .59 & .28 & .41 & .43 & .45 \\
    BEST-RQ (ViT) & .47 & \underline{.33} & \underline{.53} & .44 & .43 \\
    BEST-RQ-2 & .51 & \textbf{.41} & \textbf{.58} & \textbf{.50} & \underline{.49} \\
    \hline
  \end{tabular}
  \vspace{-0.2cm}
\end{table}

Table~\ref{tab:xares-mlp} reports linear-probing performance. Replacing the Conformer encoder with a ViT (BEST-RQ $\rightarrow$ BEST-RQ (ViT)) redistributes performance across domains: speech performance decreases while music and environmental sounds improve, yielding similar global performance. This suggests that the encoder change primarily alters inductive bias rather than overall representation quality.

The introduction of the two-step contextualize--then--predict decomposition (BEST-RQ-2) produces consistent gains over the one-stage ViT baseline across all domains, resulting in the best mean-of-means score among BEST-RQ variants while keeping inference cost unchanged. The results therefore indicate that the main performance improvement arises from the prediction decomposition rather than from the ViT encoder itself.

Among broader baselines, Whisper achieves the strongest Speech and Overall scores, which is expected given its speech-heavy training and the speech-centric composition of X-ARES~\cite{zhang2025xares}.

\subsubsection{kNN}

\begin{table}[t]
  \centering
  \setlength{\tabcolsep}{3pt}
  \caption{Condensed X-ARES kNN results ($\uparrow$), averaged by audio domain. Bold and underlined values indicate the best and second-best score, respectively, within each column across the compared models. Env.\ denotes the environmental-sound domain. MoM is the mean of the three domain means; Overall is the mean over all datasets.}
  \vspace{-0.1cm}
  \vspace{-0.1cm}
  \label{tab:xares-knn}
  \rowcolors{2}{white}{gray!13}
  \begin{tabular}{lccccc}
    \hline
    \textbf{Model} & \textbf{Speech} & \textbf{Env.} & \textbf{Music} & \textbf{MoM} & \textbf{Overall} \\
    \hline
    data2vec~\cite{baevski2022data2vec} & \textbf{.44} & .07 & .13 & .21 & .31 \\
    wav2vec~2.0~\cite{baevski2020wav2vec} & .26 & .14 & .27 & .22 & .24 \\
    Whisper~\cite{radford2023robust} & \underline{.33} & .14 & .32 & .26 & .29 \\
    Audio-JEPA & .32 & \textbf{.32} & \textbf{.58} & \textbf{.41} & \textbf{.37} \\
    BEST-RQ & .29 & .18 & .27 & .25 & .26 \\
    BEST-RQ (ViT) & .24 & .19 & .28 & .24 & .25 \\
    BEST-RQ-2 & .32 & \textbf{.32} & \underline{.49} & \underline{.38} & \underline{.35} \\
    \hline
  \end{tabular}
  \vspace{-0.2cm}
\end{table}

Table~\ref{tab:xares-knn} evaluates representation geometry using kNN classification. BEST-RQ-2 outperforms both BEST-RQ and BEST-RQ (ViT) across all aggregated metrics, indicating improved neighborhood structure beyond patch tokenization alone. Audio-JEPA remains the strongest overall non-speech method, while BEST-RQ-2 substantially improves music and environmental-sound performance relative to BEST-RQ, reinforcing the benefit of the two-step prediction design.

\subsection{XARES-LLM}
XARES-LLM evaluates a different usage regime from X-ARES: rather than probing generic representation quality, it measures how well frozen encoders function as front-end modules for large audio language models. In its current release, XARES-LLM spans 20 datasets across two tracks, covering a broad range of tasks~\cite{xares-llm}. Models pretrained with objectives aligned to text-conditioned tasks (e.g., ASR or translation) may be advantaged because their representations already emphasize factors useful for language-model training. More generally, transfer studies show representations transfer best to downstream tasks closer to the pretraining objective or data domain~\cite{yosinski2014transferable, kornblith2019better, zoph2020rethinking}. Consistent with this perspective, Whisper remains a particularly strong baseline in this setting due to its text-aligned supervision.

Table~\ref{tab:xares-llm} reports results averaged across tracks. Whisper achieves the strongest global performance, while Dasheng~\cite{dinkel2024scaling} benefits from much larger-scale self-supervised pretraining ($\sim$272.4k hours) than our AudioSet-based setup ($\sim$5.3k hours).

Within the BEST-RQ family, switching to a ViT encoder alone (BEST-RQ (ViT)) does not improve overall performance relative to BEST-RQ, reflecting a redistribution of performance across domains rather than a global gain. In contrast, BEST-RQ-2 consistently improves environmental and music performance while maintaining speech results comparable to those obtained with BEST-RQ\@. This mirrors the X-ARES observations and confirms that the contextualize--then--predict decomposition, rather than tokenization itself, is the primary source of improvement.

\begin{table}[t]
  \centering
  \setlength{\tabcolsep}{3pt}
  \caption{Condensed XARES-LLM results ($\uparrow$), averaged by audio domain across all tracks. Bold and underlined values indicate the best and second-best score, respectively, within each column across the compared models. Env.\ denotes the environmental-sound domain. MoM is the mean of the three domain means; Overall is the mean over all datasets.}
  \vspace{-0.1cm}
  \vspace{-0.1cm}
  \label{tab:xares-llm}
  \rowcolors{2}{white}{gray!13}
  \begin{tabular}{lccccc}
    \hline
    \textbf{Model} & \textbf{Speech} & \textbf{Env.} & \textbf{Music} & \textbf{MoM} & \textbf{Overall} \\
    \hline
    Dasheng~\cite{dinkel2024scaling} & \textbf{.60} & \underline{.37} & .42 & \underline{.46} & \underline{.49} \\
    Whisper~\cite{radford2023robust} & \textbf{.60} & \textbf{.41} & \underline{.58} & \textbf{.53} & \textbf{.54} \\
    Audio-JEPA & .27 & .22 & .56 & .35 & .31 \\
    BEST-RQ & .35 & .22 & .49 & .35 & .33 \\
    BEST-RQ (ViT) & .26 & .22 & .50 & .33 & .30 \\
    BEST-RQ-2 & .33 & .33 & \textbf{.59} & .41 & .38 \\
    \hline
  \end{tabular}
  \vspace{-0.2cm}
\end{table}

\section{Discussion}

Across X-ARES and XARES-LLM, replacing the Conformer in BEST-RQ with a patchified ViT primarily shifts inductive bias rather than overall transfer: speech decreases, while music and environmental sound improve, yielding similar aggregate performance. This is consistent with a tokenization trade-off where patch tokens emphasize local time--frequency texture, whereas strip-like tokens better preserve fine temporal cues important for speech.

BEST-RQ-2's main gains come from the two-step architecture design under the BEST-RQ discrete-target objective. By processing only unmasked patches in the encoder and delegating masked discrete-target prediction to a lightweight module during pretraining, BEST-RQ-2 improves transfer while keeping inference compute unchanged by discarding the predictor.

A remaining limitation is the residual speech gap relative to strip-tokenized BEST-RQ\@. Future work could explore alternative tokenization techniques or learned speech representations to improve this trade-off.

\section{Conclusion}
We introduced BEST-RQ-2, a self-supervised method for audio representation learning that retains BEST-RQ's frozen random-projection targets while adopting a two-step encoder--predictor architecture. Under a controlled pretraining budget on the same AudioSet split, BEST-RQ-2 improves transfer on music and environmental sound tasks and remains competitive on speech across both X-ARES and XARES-LLM evaluations.

Our analysis shows that patch tokenization primarily controls a token-level time--frequency trade-off, with lower performance on speech but higher performance on environmental sounds and music, whereas the encoder--predictor decomposition provides consistent gains without increasing inference compute cost. These findings highlight the importance of separating tokenization choices from prediction architecture when designing generic audio representation learners.

Future work will target improving speech-task performance while preserving gains on environmental sounds and music.

\newpage

\section{Acknowledgments}
This work was granted access to the HPC resources of IDRIS under the allocation AD011014754R2 made by GENCI.\\
Support from the ANR-3IA Artificial and Natural Intelligence Toulouse Institute ANITI (ANR-19-PI3A-0004) is gratefully acknowledged.

\section{Generative AI Use Disclosure}

We used generative AI tools to (i) proofread and improve the clarity of the writing and (ii) assist with writing and refactoring code used in the experiments. All scientific ideas, methodological choices, and the majority of the implementation and experimental work were produced and verified by the authors, who also reviewed and edited all AI-assisted outputs.

\bibliographystyle{IEEEtran}
\bibliography{mybib}

\end{document}